\documentstyle[tighten,aps,preprint,eqsecnum,epsf]{revtex}
\begin{document}
\draft
\title{Order Parameter Suppression in Double Layer Quantum Hall Ferromagnets}
\author{Yogesh N. Joglekar \thanks{Email: yojoglek@indiana.edu} \footnote{Ph: 1-812-855-7359, Fax: 1-812-855-5533}, Allan H. MacDonald}
\address{Department of Physics,\\ Indiana University,\\ Bloomington IN 47405}
\maketitle
\begin{abstract}
Double-layer quantum Hall systems at Landau level filling factor $\nu=1$ have a broken symmetry ground state with spontaneous interlayer phase coherence and a gap between symmetric and antisymmetric subbands in the absence of interlayer tunneling.  We examine the influence of quantum fluctuations on the spectral function of the symmetric Green's function, probed in optical absorption experiments. We find that as the maximum layer separation at which the $\nu=1$ quantum Hall effect occurs is approached,  absorption in the lowest Landau level grows in strength.  Detailed line shapes for this absorption are evaluated and related to features in the system's collective excitation spectrum.
\end{abstract}

\section{Introduction}
\label{sec:intro}

During recent years double quantum wells in the quantum Hall regime have been a subject of intensive study. These systems consist of two 2-dimensional electron layers in a perpendicular magnetic field with a distance $d$ ($ d\approx$ 100 \AA)  between layers comparable to the typical distance between electrons in the same layer.  When the magnetic field is strong enough to accommodate all electrons in the lowest Landau level (LLL), interactions between the electrons largely determine the properties of the system. Even when the spin degree of freedom can be ignored because of complete spin alignment, the system exhibits a rich variety of phases~\cite{kmoon} associated with the layer index degree-of-freedom and dependent on the difference between interlayer and the intralayer Coulomb interactions.  These states are referred to as quantum Hall ferromagnets.  In particular, even in the absence of a finite tunneling amplitude, there is \emph{spontaneous} interlayer phase coherence, which lifts the degeneracy between  single-particle symmetric states which are occupied and antisymmetric states which are empty~\cite{brey,wen,ezawa}. In a mean-field theory this splitting blocks optical absorption in the lowest Landau level at T=0.  Absorption is permitted only because of quantum fluctuations, making this probe particularly important. In this paper we present a theory of quantum fluctuations and optical absorption in double-layer quantum Hall ferromagnets.

\section{Formalism}
\label{sec:formal}

In the following, we discuss the nature of the many-body ground state wavefunction for such a system in a mean field approximation and systematically improve upon it by including the effect of quantum fluctuations.   We present some numerical results and briefly discuss their experimental implications.  Let us consider a system at filling factor $\nu$=1, neglect the spin dynamics and use the lowest Landau level  approximation.  It is convenient to describe the layer index degree-of-freedom by a  `pseudospin', where the symmetric state corresponds to `pseudospin up' ($|\uparrow\rangle$) and the antisymmetric state is `pseudospin down' ($|\downarrow\rangle$). Then the interaction between the electrons is a sum of two potentials: a term $V_0=(V_A+V_E)/2$, which conserves  pseudospin and a term $V_x=(V_A-V_E)/2$, which reverses the pseudospins of the interacting electrons. ($V_A$ and $V_E$ are the intralayer and interlayer Coulomb interactions respectively.) We expect the mean-field ground state to be fully pseudospin polarized, with all electrons occupying the symmetric single-particle orbitals.  Since the $V_x$ term flips pseudospins, however, it is clear that the \emph{exact} ground state must have an indefinite pseudospin polarization.  Hence \emph{ even at zero temperature}, there must be some mixing of reversed pseudospins in the true ground state.  We calculate this mixing by considering the scattering of electrons off of virtually excited collective excitations, pseudospin-waves. The finite temperature expression for the symmetric-state self energy is given by~\cite{yogesh}

\begin{equation}
\label{eqn:segeneral}
\Sigma_S(i\omega_n)=-\frac{1}{\beta}\sum_{i\Omega}\sum_{a=S,AS}{\cal G}^{MF}_a(i\omega_n-i\Omega)M^{-1}_{Sa,aS}(i\Omega),
\end{equation}
where ${\cal G}_a^{MF}(i\omega_n)=(-i\omega_n+\xi_a)^{-1}$ is the mean-field Matsubara Green's function and $M^{-1}$ is the pseudospin-wave propagator matrix. At zero temperature the symmetric-state self energy becomes

\begin{equation}
\label{eqn:sezeroT}  
\Sigma_S(i\omega_n)=\frac{2\pi l^2}{A}\sum_{\vec{q}}\frac{(E_{sw}(\vec{q})+\Delta)^2}{2E_{sw}(\vec{q})}\frac{\epsilon(\vec{q})-E_{sw}(\vec{q})}{i\omega_n-\xi_{AS}-E_{sw}(\vec{q})}.
\end{equation}
Here $E_{sw}(\vec{q})$ is the pseudospin-wave energy\cite{fertig}, $\xi_{AS}=-\xi_S >0$ is the mean-field energy of the antisymmetric state and $\Delta=2\xi_{AS}$ is the interaction enhanced quasiparticle level splitting between the symmetric and the antisymmetric state energies.

For models with delta-function electron-electron interactions, like the Hubbard model which is frequently used for theories of itinerant electron magnetism,  self-energy expressions of this form are most efficiently done by using the Hubbard-Stratonovich transformation~\cite{negele} after formally expresssing the  electron-electron interaction as an exchange interaction favoring parallel spin alignment.  For double layer systems, however, this transformation is not possible and both Hartree and exchange fluctuations are important in the collective excitation spectrum and the fluctuation physics.  To make progress we have derived the above results using a modified version of the Hubbard-Stratonovich transformation~\cite{kerman} designed to cope with this difficulty, which is almost always present in realistic models. 

\begin{figure}[h]
\begin{center}
\epsfxsize=3in
\epsffile{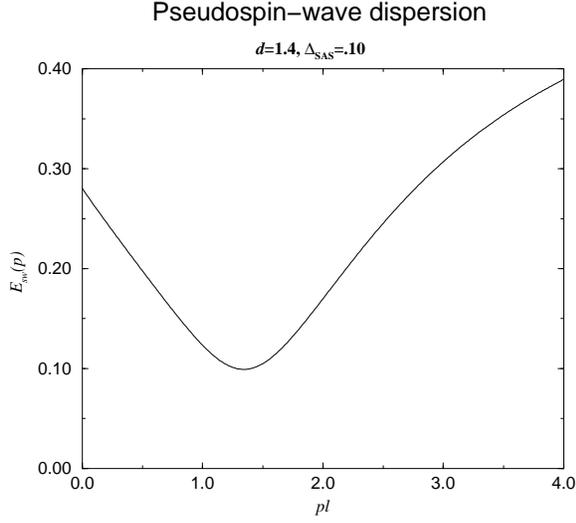}
\vspace{.5cm}
\caption{Collective pseudospin-wave excitation dispersion relation for $d=1.4l$ and single-particle tunneling gap $\Delta_{SAS}=0.1 e^2/\epsilon l$.  Note the minimum at finite wavevector. }
\label{fig:dispersion}
\end{center}
\end{figure}

In magnetic systems, spin wave energies usually increase  monotonically with momentum so that low-energy physics is long-wavelength physics, which in turn can be described by a continuum (effective) field theory.  In contrast, for a double layer system, the pseudospin collective mode energies usually have a minimum at $pl\approx 1$ where $l$ is the magnetic length ~\cite{fertig} and so an effective field theory discription is not useful. (See fig.~\ref{fig:dispersion}). We also note that when the distance between the layers vanishes, the pseudospin flipping interaction vanishes, and then the mean-field approximation for the ground state is exact. This is reflected in the vanishing zero temperature self energy expression (~\ref{eqn:sezeroT}) when $d=0$.   

We analytically continue the above self-energy expression to real frequencies and solve the Dyson's equation, $\omega-\xi_S=\Sigma_S(\omega+i\eta)$, numerically to obtain the spectral function for the symmetric state. In the absence of the fluctuation self-energy correction, the thermal Green's function is given by ${\cal G}_S(i\omega_n)={\cal G}^{MF}_S(i\omega_n)$. This corresponds to a spectral function $A_S(\omega)=\delta(\omega-\xi_S)$;  all the spectral weight is in the delta function at the negative energy (occupied) symmetric state quasiparticle pole. When the self energy correction is included, the spectral weight at the symmetric state quasiparticle pole reduces to $z_S=(1-\partial\Sigma_S/\partial\omega|_{\omega^*})^{-1}$ where $\omega^*$ satisfies the Dyson's equation. The remaining spectral weight is distributed in a continuum piece at positive (unoccupied) energies where the self-energy (~\ref{eqn:sezeroT}) has a branch cut, \emph{i.e.} in the interval $\xi_{AS}+E_{sw}^{min}\leq \omega\leq \xi_{AS}+\Delta$.   If excitonic interactions can be neglected the lowest-Landau-level contribution to the optical absorption spectrum is given by the positive energy part of $A_S(\omega)$.  One index of the amount of spectral weight at positive frequencies is the suppression of the ground-state pseudospin polarization,

\begin{equation}
\label{eqn:mag}
m(T=0)= z_S = \int^{\infty}_{-\infty}\frac{d\omega}{2\pi}n_F(\omega)\left[A_S(\omega)-A_{AS}(\omega)\right],
\end{equation}
from it's mean-field value.

\section{Results}

\begin{figure}[h]
\begin{center}
\epsfxsize=3in
\epsffile{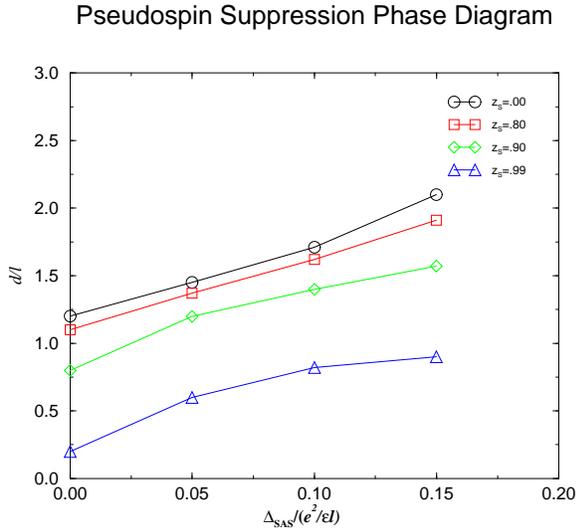}
\vspace{.5cm}
\caption{Ground state pseudospin polarization contours. In this theory the $z_S=0$ line is coincident with the stability boundary of the quantum Hall state.  Polarization suppression and positive energy symmetric spectral weight grow rapidly as the stability boundary is approached. }
\label{fig:phase}
\end{center}
\end{figure}

\begin{figure}[h]
\begin{center}
\epsfxsize=3in
\epsffile{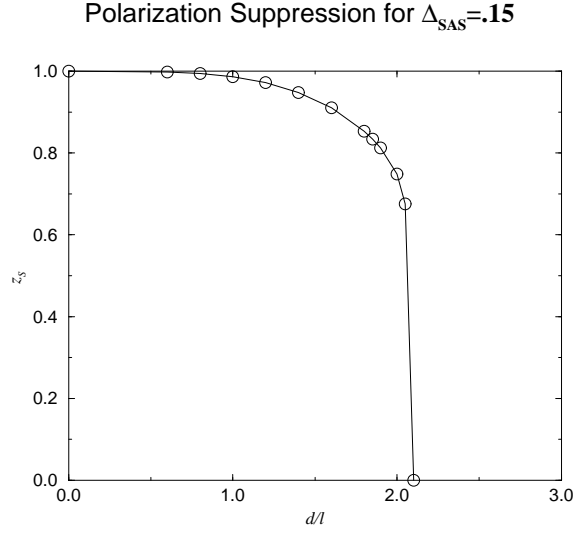}
\vspace{.5cm}
\caption{Ground state pseudospin polarization \emph{vs.} layer separation for $\Delta_{SAS} =0.15e^2/\epsilon l$.}
\label{fig:polarization}
\end{center}
\end{figure}

\begin{figure}[h]
\begin{center}
\epsfxsize=3in
\epsffile{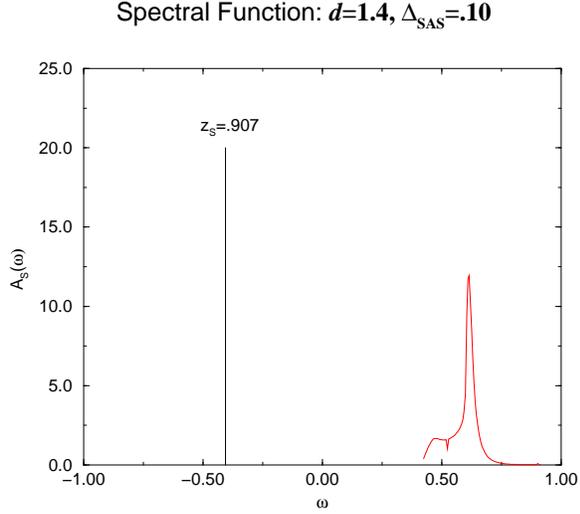}
\vspace{.5cm}
\caption{Symmetric spectral function for $d=1.4l$ and $\Delta_{SAS}=0.1 e^2/\epsilon l$. The zero of energy is at the chemical potential, and energy is measured in units of $e^2/\epsilon l^2$. The positive energy spectral function is proportional to the lowest Landau level contribution to the optical absorption spectrum, neglecting excitonic corrections. }
\label{fig:spectralfn}
\end{center}
\end{figure}

We have calculated the pseudospin polarization for various values of the two experimentally controlable parameters in these systems, namely the distance between the layers $d$ and the tunneling amplitude $\Delta_{SAS}$. (We have neglected the finite thickness of the electron layers to simplify the calculations).  A  phase diagram showing curves of equal polarization is plotted in fig.~\ref{fig:phase}. The line corresponding to $z_S=0$ is the same as the phase boundary between QHE/NO-QHE regions~\cite{ahmtheory,murphy}.  As we approach the phase boundary, the minimum of the pseudospin-wave energy, $E_{sw}^{min}$ (occuring at a finite wavevector) approaches zero, leading to the instability which destroys the pseudospin polarized state and also the quantum Hall effect. The polarization drops to zero rather sharply as a function of $d$ for a given $\Delta_{SAS}$ (see fig.~\ref{fig:polarization}) since $E_{sw}^{min}$ vanishes rapidly with  layer separation. The spectral function for $d=1.4$ and $\Delta_{SAS}=.10$ is shown in fig.~\ref{fig:spectralfn}. We see that in this case, the pseudospin polarization is reduced by 10\% from its maximum value of 1. The nonvanishing spectral function at positive energies reflects the possibility of adding a symmetric electron as an antisymmetric state quasiparticle by destroying a  pseudospin-wave present in the ground state.

We stress the important role of special properties of the lowest Landau level single-particle states and the absence of band structure in simplifying the calculation described here.  The  presence, at $\nu=1$ and low temperatures, of positive energy  symmetric state spectral weight  has been detected recently by Manfra and Goldberg ~\cite{manfra} in a sample which is close to the quantum Hall boundary. Above theoretical results on the one-electron Green's function of double-layer systems provide a starting point for interpreting optical absoprtion experiments like those reported in Ref.~\cite{manfra}. If the excitonic effects can be ignored, the optical absorption spectrum is proportional to the portion of the symmetric spectral weight at energies above the Fermi energy. For $\nu=1$, the contribution from the lowest Landau level vanishes since the symmetric orbital is occupied. Existing experiments do show a clear evidence of the absorption due to quantum fluctuations expected on the basis of these calculations. Extending the above numerical calculations to take into account the finite thickness of the layers is possible without much difficulty. It will be interesting to see if the predictions for positive energy spectral weight made by this theory are in agreement with experiments as they are refined and as the theory is refined by accounting for excitonic effects.

\section*{acknowledgement}
This work was supported by NSF grant no. DMR9714055. 

\begin{thebibliography}{99}
\bibitem{brey}L. Brey, Phys. Rev. Lett. {\bfseries 65}, 903 (1990)
\bibitem{wen}X. G. Wen and A. Zee, Phys. Rev. Lett. {\bfseries 69}, 1811 (1992).
\bibitem{ezawa}Z.F. Ezawa and A. Iwazaki, Int. J. Mod. Phys. B {\bfseries 19}, 3205 (1992).
\bibitem{kmoon} K. Moon {\it et al.}, Phys. Rev. B {\bfseries 51}, 5138 (1995).
\bibitem{yogesh} Y. N. Joglekar and A. H. MacDonald, in preparation (1999).
\bibitem{fertig} H. A. Fertig, Phys. Rev. B {\bfseries 40}, 1087 (1989).
\bibitem{negele} J. W. Negele and H. Orland, {\it Quantum Many-Particle Systems} (Addison-Wesley, New York, 1988).
\bibitem{kerman} A. K. Kerman, S. Levit and T. Troudet, Ann. Phys. {\bfseries 148}, 436 (1983).
\bibitem{ahmtheory}  A.H. MacDonald, P.M. Platzman, and G.S. Boebinger,  Phys. Rev. Lett. {\bf 65}, 775 (1990).
\bibitem{murphy} S. Q. Murphy, J. P. Eisenstein, G. S. Boebinger, L. N. Pfeiffer and K. West, Phys. Rev. Lett. {\bfseries 72}, 728 (1994).
\bibitem{manfra} M. J. Manfra {\it et al.},  cond-mat/9809373.
\end{thebibliography}
\end{document}